\documentclass[fleqn]{article}
\usepackage[widespace]{fourier}
\usepackage[psamsfonts]{eucal}         
\usepackage{graphicx}
\usepackage{cite}
\usepackage[cmex10]{amsmath}
\usepackage{amssymb}
\usepackage{bm}
\usepackage{psfrag}
\usepackage[font=small]{caption}
\usepackage[usenames,dvipsnames]{color}
\renewcommand{\i}{\mathrm{i}}
\newcommand{\e}{\mathrm{e}}
\newcommand{\half}{{\textstyle \frac{1}{2}}}
\renewcommand{\vec}[1]{\boldsymbol{#1}}
\newcommand{\SE}{Schr\"o\-ding\-er equation}
\newcommand{\SSE}{Schr\"o\-ding\-er Equation}
\newcommand{\fl}{}
\newcommand{\rmd}{d}
\newcommand{\rme}{\e}
\newcommand{\rmi}{\i}
\newcommand{\x}{\boldsymbol{x}}
\newcommand{\q}{\boldsymbol{q}}
\newcommand{\p}{\boldsymbol{p}}
\newcommand{\kbf}{\boldsymbol{k}}
\newcommand{\psihat}{\hat{\psi}}
\newcommand{\phihat}{\hat{\phi}}
\newcommand{\ofx}{(\x)}
\newcommand{\etal}{\emph{et al}}

\newcommand\BEC{Bose--Einstein condensate}
\newcommand{\BBECn}{Bose--Einstein Condensation}
\newcommand\BECn{Bose--Einstein condensation}
\newcommand{\GPE}{Gross--Pitaevskii equation}

\newcommand{\paperii}{\emph{Paper II}}
\newcommand{\paperiii}{\emph{Paper III}}

\title{Effective Field Theory for Atom-Molecule Systems I:
Formulation of Effective Field Theory}
\author{Catarina E Sahlberg and C W Gardiner
\\[5mm] \emph{Jack Dodd Centre for Quantum Technology,} 
\\ \emph{Department of Physics, University of Otago,}
\\ \emph{ Dunedin, New Zealand}}
\begin{document}
\maketitle

\begin{abstract}
We present a model of a coupled bosonic atom-molecule system, using
the recently developed c-field methods as the basis in our formalism.
We derive expressions for the s-wave scattering length and binding
energy within this formalism, and by relating these to the
corresponding experimental parameters, we can accurately determine
the phenomenological parameters in our system.
\end{abstract}

\section{Introduction}
C-field methods have become an indispensable tool in the quantitative
description of many aspects of \BECn\ physics \cite{blakie2008},
providing a description of the dynamics of highly degenerate Bosonic
gases, which incorporates quantum mechanics correctly, and is accurate
provided the density of the Bose gas is sufficiently high.

The c-field method is based on the use of a Wigner function
representation, in which a truncation approximation is made, and this
is valid when the density of the \BEC\ is large---for details the
reader is referred to \cite{blakie2008}.  The mathematical formulation
of the c-field method which results from this is superficially very
similar to that provided by the \GPE. The principal feature in
addition to the \GPE\ is provided by the inclusion of a stochastic
representation of quantum fluctuations in the initial conditions.  In
the high density limit the method provides a treatment in which
quantum and thermal phenomena are correctly accounted for.  The
predictions of c-field calculations can be dramatically different from
those of the \GPE, as was shown in the treatment of colliding
condensates \cite{Norrie2005a,Norrie2006,Norrie2006b}, in which the
c-field method was first introduced.

The quantum fluctuations correspond to half a quantum of noise added
to each degree of freedom; to avoid ultraviolet divergence it is
therefore necessary to restrict the number of modes used.  In practice
this is done by means of a projection of the equations of motion into
a subspace with a maximum momentum, usually called $ \hbar\Lambda$.
Such a cutoff is necessary for two other reasons:
\begin{enumerate}
\item Numerical computations must always be restricted to a finite
number of modes.  This is most commonly provided by the spatial grid,
although, as shown in \cite{blakie2008}, this must be implemented as a
projector into the relevant mode subspace if aliasing is to be
avoided.  This issue is not so important for simple \GPE\ simulations,
since the occupations of modes at risk of being aliased are usually
small, but the in the c-field method \emph{every} mode has at least a
half quantum of occupation, and aliasing is definitely an issue of
concern.

\item 
As in the Gross-Pitaevskii equation, the interactions between the
condensed particles in the c-field formalism are represented by a
localized contact potential.  This pseudopotential method implies the
use of a cutoff, as has been argued by Braaten and Nieto
\cite{braaten1997}, who introduced the requirement on the cutoff $
\Lambda$ and the interparticle scattering length $ a_s$
\begin{eqnarray}\label{cutoff1}
\Lambda a_s \ll 1,
\end{eqnarray}
for the validity of a pseudopotential method.

As implied by our use of the same notation $\Lambda$, the
pseudopotential cutoff and that required by the numerical simulations
are in practice essentially identical.  (In principle there is a
difference, discussed in the appendix.)  In practice the choice of
cutoff mandated by the numerical algorithms used satisfies this
criterion with a satisfactory margin of safety; for example in the
simulations done by Norrie \etal\
\cite{Norrie2005a,Norrie2006,Norrie2006b} the cutoff used satisfied $
\Lambda a_s\approx 0.1 $.
\end{enumerate} 

Apart from the solitary case of hydrogen condensates, in all \BEC\
experiments 
the scattering length $ a_s$ used to describe the interactions arises
because 
of the existence of a weakly bound state of two atoms. The use of
Feshbach 
resonances can give very large scattering lengths, consequently reducing the 
binding energy of this state to a very small value.  This can
introduce quite 
slow time scales, which makes it wise to investigate the possible
influence of 
molecular dynamics on \BECn\ phenomena.  

The c-field method cannot 
be directly applied to any explicit description of the relevant
molecular 
physics, since binding can only be described by an attractive
potential, for 
which the \GPE\ and similar c-field equations have no stable
solutions, and 
certainly do not produce molecules.  In this and subsequent papers,
we want to 
combine the ideas of c-fields with those field theory methods which
use an explicit 
``molecule field'', as originally introduced in 
\cite{Heinzen2000a,holland2001}.  These kinds of models 
are purely phenomenological descriptions of the physics, whose
parameters must 
be determined to reproduce the correct experimentally measurable
quantities. 

The two relevant parameters for the molecular field model are the 
\emph{binding energy} of the weakly bound state, and the \emph{s-wave 
scattering length}, which can be measured. We will therefore
develop a method for relating them to the phenomenological parameters
in our formalism. The effective range, we will argue, is not a useful
parameter for characterizing the physics. 

The formalism outlined in this paper is used in \paperii\
\cite{sahlberg2}, where we implement the mean field theory in
the Thomas-Fermi approximation, and the Bogoliubov theory for this
model of a coupled atom-molecule system and use the latter to
investigate the excitation spectrum for Bragg scattering from a
uniform condensate. In \paperiii\ \cite{sahlberg3} we implement this
formalism numerically, performing full simulations of Bragg scattering
from a trapped Bose-Einstein condensate, as in the recent experiment
by \cite{papp2008}. The results from the Bogoliubov calculations and
the full simulations both show that the 
measured effects on Bragg scattering as the scattering length
increases are
well described by our coupled atom-molecule formalism.


\section{Effective Hamiltonian Method for \BBECn}
It is normal when describing \BECn\ to use an approximate Hamiltonian
for the system of ultra-cold atoms that gives a good description of
only
the long-wavelength behaviour of the system, which is all that is
relevant for the observable phys\-ics.  The relevant methods are called
``pseudopotential methods'' or ``effective field theory methods'', and
there 
is a long history associated with the  
various formulations of these methods 
\cite{Breit1947a,Blatt1952a,Huang1957a,braaten1997}.  In all of these 
formulations, the underlying philosophy 
is to find an approximate description of the the physics of very low
energy 
particles. The definition of ``low energy'' in practice is that the
scattering 
amplitude for such energies does not differ significantly from its
value at 
zero energy.  

It is best to formulate the concepts required  precisely, and in a
form adapted 
to the study of a trapped ultra-cold gas.  In such a system all of
the relevant 
physics involves particles with a finite small  momentum---this means
that we
can restrict the description to particles with momentum less than  a
fixed 
cutoff 
value $ \hbar \Lambda$.  Consequently, the maximum
\emph{relative} 
momentum is  $ 2\hbar \Lambda$, and for a given centre of mass
momentum 
$ \vec Q$ of any pair of colliding particles the allowable values of
relative 
momentum, $ \vec p$, satisfy
$ |\vec p +\half  \vec Q | \le  \Lambda$.  Thus the description of
scattering 
of any pair of particles depends on their centre of mass momentum.
If the 
possible values of $ \vec Q$ are themselves rather small in
comparison with
$ \hbar \Lambda$, this dependence on the centre of mass momentum is
not very 
important, and it is normally ignored.

The appropriate quantum field theory description in the case that the
centre 
of 
mass dependence can be neglected is essentially that of Braaten and
Nieto 
\cite{braaten1997}, and can be characterized as:
\begin{enumerate}
\item  The system of atoms is described by a quantum field operator
 $\psihat$, defined on the low energy subspace
specified by a momentum space cutoff $\Lambda$
\begin{eqnarray}
\psihat\ofx = \frac{1}{(2\pi)^{3/2}}\int_0^\Lambda d\kbf\,{a_{\kbf}
\rme^{\rmi\kbf\cdot\x}}.
\end{eqnarray} 
\item The Hamiltonian for a trapped
Bosonic gas of interacting particles is given by
\begin{eqnarray}
\label{eq:H_basic}
H = \int{\rmd\x\,\,
\left\{\psihat^\dagger\ofx\left(-\frac{\hbar^2\nabla^2}{2m}+V_a\ofx\right)
\psihat\ofx
+
\frac{U}{2}\psihat^\dagger\ofx\psihat^\dagger\ofx\psihat\ofx\psihat\ofx\right
\}},
\end{eqnarray}
where $V_a$ is the trapping potential.
\item The quantity $U$ is the inter-atomic
scattering strength, and is related to the s-wave scattering length
$a_s$ by
\begin{eqnarray}
U = \frac{4\pi\hbar^2 a_s}{m(1 - 2\Lambda a_s/\pi)}.
\end{eqnarray}
\end{enumerate}
For the method to be useful it is necessary that $ 2\Lambda a_s/\pi
\ll 1 $, which will mean that the predictions of this Hamiltonian are
independent of the cutoff, as long as this is not too large.%
\footnote{In fact, there is no reason to believe that the
Hamiltonian (\ref{eq:H_basic}) is valid unless this condtion is
satisfied.}
In this case, at sufficiently low temperatures, the
condensate wavefunction $ \Psi(\vec x,t)$ is accurately described by
the \GPE\
\begin{eqnarray}\label{GPE}
\i\hbar{\partial \Psi\over\partial t} &=&
{}-{\hbar^2\nabla^2\Psi\over 2m} +V(\vec x)\Psi 
+{4\pi\hbar^2a_s\over m}|\Psi |^2\Psi  .
\end{eqnarray}
Experimentally, this is a well-verified equation.  

If $ 2\Lambda a_s/\pi $ is not sufficiently small, the simple
relationship between the Hamiltonian and the \GPE\ disappears.  The
derivation of the \GPE\ from the Hamiltonian involves higher order
terms in perturbation theory, which the choice of a sufficiently small
$ \Lambda $ implicitly sums.

It is important to emphasize that, although $ \Lambda$ is often called
an ``ultraviolet cutoff'', unlike such cutoffs in quantum
electrodynamics, it has a finite value, and this value must be
\emph{small} for the pseudopotential Hamiltonian to be valid.
Furthermore, in c-field methods, all states have at least half a
quantum of occupation, so that all momenta up to $ \hbar\Lambda$
participate in calculations, and the corrections discussed in the
appendix may be relevant.

\section{An Effective Field Method for Molecules}
Let us now introduce molecules into the c-field 
formalism.  Since the \GPE\ cannot produce bound states, it is clear
that in 
some sense these molecules must be introduced ``by hand''.  

\subsection{The Molecular-Field Hamiltonian}
A method used by several other groups
\cite{Drummond1998a,timmermans1999,Heinzen2000a,holland2001,Duine2003a},
is to
add an additional field $\phihat$ corresponding to a molecule state,
giving the Hamiltonian:
\begin{eqnarray}
\label{eq:H}
\fl
\hat{H} = \int{\rmd\x\,\,
\left\{\psihat^\dagger\ofx\left(-\frac{\hbar^2\nabla^2}{2m}+V_a\ofx\right)
\psihat\ofx
+\phihat^\dagger\ofx\left(-\frac{\hbar^2\nabla^2}{4m}+V_m\ofx+\varepsilon
\right)
\phihat\ofx
+ \right.} \nonumber \\
\left.\frac{U_{aa}}{2}\psihat^\dagger\ofx\psihat^\dagger\ofx\psihat\ofx\psihat
\ofx
+
{U_{am}}\psihat^\dagger\ofx\phihat^\dagger\ofx\phihat\ofx\psihat\ofx
+ \right. \nonumber \\
\left.\frac{U_{mm}}{2}\phihat^\dagger\ofx\phihat^\dagger\ofx\phihat\ofx\phihat
\ofx
+ \frac{g}{2}\left(\phihat^\dagger\ofx\psihat\ofx\psihat\ofx +
\psihat^\dagger\ofx\psihat^\dagger\ofx\phihat\ofx\right) \right\}.
\end{eqnarray}
Here the parameters have the interpretations
\begin{enumerate}
\item
$U_{aa}$ is the background atom interaction strength, leading to the
concept of a background scattering length $ a_{bg}\equiv
mU_{aa}/4\pi\hbar^2.$

\item  
$V_a$ and $V_m$ are the external trapping potential for the atoms and
the molecules respectively.  If the magnetic moment of
the weakly bound molecule is twice that of the atom, then 
$V_m=2V_a$. However, this is not mandatory.

\item
$\varepsilon$ is an energy offset term, which allows for a finite 
binding energy of the molecule.

\item The coupling parameter $g$ describes strength of coupling 
of the process by which a molecule is formed from two atoms.

\item
The terms with factors $U_{am}$
and $U_{mm}$ correspond to atom-molecule and molecule-molecule
scattering respectively. Since the atom field is usually much
larger than the molecule field in the situations we shall consider,
these terms are in most cases negligible. 
\end{enumerate}
\subsubsection{Phenomenology}
The Hamiltonian (\ref{eq:H}), it must be emphasized, provides only a 
phenomenological description of the physics.  In particular, the
energy offset 
$ \varepsilon$ can be viewed as a representation of the effect of a
Feshbach 
resonance, and is related to the binding energy of the molecule.
However, the 
actual binding energy must be determined by solving the appropriate
\SE, and it 
will depend on the other parameters.  Furthermore, for each value of
$ \varepsilon$, the other parameters, and in particular $ g$, may be
different, 
and indeed, we will find that the values of $ g$ and $ \varepsilon$
required to 
fit the measured scattering and binding properties are strongly
interdependent.
These dependencies in practice simply mean that when setting up a
c-field 
simulation, one must choose the parameters appropriate to the
experimental 
system under investigation.

In summary, the sole function of the Hamiltonian (\ref{eq:H}) is to
provide a 
practical method of implementing c-field theory for molecules in a
way that is 
consistent with measured properties of the atom-molecule system. The
basic 
theory to which it is an approximation is a Hamiltonian involving
only atoms 
interacting through an appropriate interatomic potential 
$ u(\vec x - \vec x')$. The molecular field method is only necessary
in this paper because 
the ``exact'' Hamiltonian cannot be represented inside a c-field
theory.


\subsubsection {Momentum Cutoffs} 
In this case of a coupled atom and molecule system there
will be momentum cutoffs for both the molecular and the atomic fields,
and these can be expressed in terms of projectors $\mathcal{P}_a$ and
$\mathcal{P}_m$ that project the wavefunctions onto the low energy
subspace below the cutoff.  Because a molecule is formed from two
atoms, the interactions will only make sense if the molecule cutoff is
twice that of the atom.  Thus, if the atom cutoff is $\Lambda$, 
the projectors can be defined as
\begin{eqnarray}
\label{eq:Pa}
\mathcal{P}_a(\kbf) &= \Theta(\Lambda - |\kbf|),\\
\label{eq:Pm}
\mathcal{P}_m(\kbf) &= \Theta(2\Lambda - |\kbf|),
\end{eqnarray}
where $\Theta$ is the Heaviside step function. 

In (\ref{eq:Pa},\ref{eq:Pm}) we have assumed isotropic cutoffs, but
this is not necessarily the case for all systems. Indeed, as we shall
see in \paperiii\, in experimentally realistic systems the cutoff can
be highly anisotropic. However, for the work in this paper, the exact
properties of the cutoff are not relevant, and we therefore assume
that it is isotropic.

\section{Determination of Parameters}

In order to determine the relationship between this formalism and reality, the parameters and
the fields have to be related to physically observable quantities.  
To do this, we need to compute 
\begin{enumerate}
\item The scattering amplitude,
\item The binding energy,
\item The bound state wavefunction.
\end{enumerate}

\subsection{\SSE\ in the 2-Atom : 1-Molecule Sector}%
\label{sec:2atom1mol}%
If $|E\rangle$ is a state in 2-atom : 1-molecule sector, it has
a two-component  wavefunction, defined by the 2-atom amplitude
\begin{eqnarray}\label{wvfn1}
\psi(\x_1,\x_2) \equiv
\langle 0|\psihat(\x_1)\psihat(\x_2)|E\rangle ,
\end{eqnarray}
and the 1-molecule amplitude
\begin{eqnarray}\label{wvfn2}
\phi(\x)  \equiv\langle 0 |\phihat(\x)|E\rangle  ,
\end{eqnarray}
where $|0\rangle$ is the vacuum state. The normalization of the wavefunctions is given by the condition
\begin{eqnarray}
\int{|\psi(\x_1,\x_2)|^2 \rmd\x\,_1\rmd\x\,_2} +2\int{|\phi\ofx|^2\rmd\x\,}
=1.
\end{eqnarray}
We write the  Schr\"odinger equation for the atom and
molecule
wavefunctions corresponding to the Hamiltonian (\ref{eq:H}) as
\begin{eqnarray}
E\psi(\x_1,\x_2) &=&-
\frac{\hbar^2\left(\nabla^2_1+\nabla^2_2\right)}{2m}\psi(\x_1,\x_2)
+ \delta(\x_1-\x_2)\left(U_{aa}\psi(\x_1,\x_2)+g\phi(\x_1)\right),
\\ E\phi(\x) &=&
-\frac{\hbar^2\nabla^2}{4m}\phi(\x)
+\varepsilon\phi(\x)
 +\frac{g}{2}\psi(\x,\x).
\end{eqnarray}
The equivalent  Shr\"odinger equations for the momentum space
wavefunctions $\tilde{\psi}$ and $\tilde{\phi}$ take the form
\begin{eqnarray}
\left(E-\frac{\hbar^2}{2m}(p_1^2+p_2^2)\right)\tilde{\psi}(\p_1,\p_2)
&=&
\frac{U_{aa}}{(2\pi)^3}\int^\Lambda_0{\rmd\q\,_1\,\int^\Lambda_0{\rmd\q\,_2\,
\delta(\q_1+\q_2-\p_1-\p_2)\tilde{\psi}(\q_1,\q_2)}}
\nonumber 
\\ && \qquad\qquad\qquad\qquad\qquad\qquad\qquad
+ g\tilde{\phi}(\p_1+\p_2),
\\
\left(E - \varepsilon - \frac{\hbar^2p^2}{4m}\right)\tilde{\phi}(\p)
&=&
\frac{g}{2(2\pi)^{3}}\int^\Lambda_0{\rmd\q\,_1\,\int^\Lambda_0{\rmd\q\,_2\,\,
\delta(\q_1+\q_2-\p)\tilde{\psi}(\q_1,\q_2)}}. \nonumber\\
\end{eqnarray}
We transform the coordinate system to one of center-of-mass and relative
momenta by  the substitutions $\boldsymbol{P} =\p_1+\p_2$,
$\boldsymbol{P}'=\q_1+\q_2$, $\kbf =(\p_1-\p_2)/2$ and $\kbf'
=(\q_1-\q_2)/2$. In the  centre of mass frame where
$\boldsymbol{P}=0$ the equations take the form
\begin{eqnarray}\label{eq:atomyamaguchi}
\left(E -\frac{\hbar^2k^2}{m}\right)\tilde{\psi}(\kbf) &=&
\frac{U_{aa}}{(2\pi)^{3}}\int_0^\Lambda{\rmd\kbf'\,\tilde{\psi}(\kbf')}+g
\tilde{\phi}(0),
\\
\label{eq:molyamaguchi}
\qquad\left(E - \varepsilon\right)\tilde{\phi}(0)
&=&\frac{g}{2(2\pi)^{3}}\int_0^\Lambda{\rmd\kbf'\,\tilde{\psi}(\kbf')},
\end{eqnarray}
where we have redefined $\tilde{\psi}$ as the one component momentum
space wavefunction for the atomic field.

The projectors (\ref{eq:Pa},\ref{eq:Pm}) define a preferred frame; in
a frame 
in which $ \vec P \ne 0$, the range of the momentum integrals becomes 
$ |\vec k +\half\vec P| \le \Lambda$. Since the most important
interactions in 
a \BEC\ do correspond to very small centre of mass momentum (see
appendix), this is
not a 
great problem.


\subsubsection{The Yamaguchi Equation}
Substituting the molecule function (\ref{eq:molyamaguchi}) into the atom equation (\ref{eq:atomyamaguchi}), we obtain
\begin{eqnarray}
\label{eq:yamaguchi}
\left(\frac{Em}{\hbar^2} -k^2\right)\tilde{\psi}(\kbf)
=\lambda(E)\int_0^\Lambda{\rmd\kbf'\,\tilde{\psi}(\kbf')},
\end{eqnarray}
where 
\begin{eqnarray}
\label{eq:yamaguchi_coeff}
\lambda(E) =
\frac{m}{(2\pi)^{3}\hbar^2}\left(U_{aa}+\frac{g^2/2}{E-\varepsilon}\right).
\end{eqnarray}
Equation (\ref{eq:yamaguchi}) is similar in form to Yamaguchi's
separable potential equation \cite{yamaguchi1954}, with the important
difference that our expression has an explicit dependence on the
energy eigenvalue $E$ in the $\lambda$ factor. However, the method of
solution 
is essentially unaltered.


\subsubsection{Scattering state solution}
A scattering state will be characterized by a positive energy
eigenvalue, so we make the substitution $E\rightarrow\hbar^2K^2/m$ in
(\ref{eq:yamaguchi}), which can now be written as
\begin{eqnarray}
\label{eq:scatteringEqn}
\tilde\psi(\kbf) =\delta(\boldsymbol{K} -\kbf)
+\frac{\lambda_K}{K^2-k^2+i\eta}\int_0^\Lambda{\rmd\kbf'\,\tilde{\psi}(\kbf')},
\end{eqnarray}
where we have defined $\lambda_K \equiv \lambda(\hbar^2K^2/m)$.
Integrating over $\kbf$ on both sides gives the result
\begin{eqnarray}
\label{eq:scatteringPsi}
\int_0^\Lambda{\rmd\kbf\,\tilde{\psi}(\kbf)} &=&  \left(1-
\lambda_K\int_0^\Lambda{\rmd\kbf\,\frac{1}{K^2-k^2+\i\eta}}\right)^{-1}.
\end{eqnarray}
\begin{description}
\item[{Small $\vec K$ Approximation}]
The integral on the right hand side can be evaluated exactly, and for
$K\ll \Lambda$ approximated thus:
\begin{eqnarray}
\label{eq:integral}
\int_0^\Lambda{\rmd\kbf\,\frac{1}{K^2-k^2+\i\eta}} = -4\pi\Lambda -
2\i\pi^2 K +2\pi K\ln\left( \frac{\Lambda + K}{\Lambda - K}\right)
\approx -4\pi\Lambda - 2\i\pi^2 K + 4\pi \frac{K^2}{\Lambda}.
\end{eqnarray}
The atom wavefunction for the scattering state can therefore be
written as
\begin{eqnarray}
\label{eq:psi_k}
\tilde\psi(\kbf) =\delta(\boldsymbol{K} -\kbf)
-\frac{1}{(-1/\lambda_K-4\pi \Lambda - 2\i\pi^2K +4\pi K^2/\Lambda)}
\frac{1}{(K^2-k^2+\i\eta)}.
\end{eqnarray}
Similarly, the solution for the molecular wavefunction is given by
\begin{eqnarray}
\tilde\phi(0) &=&
\frac{g/2}{\hbar^2K^2/m-\varepsilon}
\int_0^\Lambda{\rmd\kbf'\,\tilde{\psi}(\kbf')}= \nonumber \\
&=&
\frac{g/2}{(\hbar^2K^2/m-\varepsilon)}
\frac{1}{\left(-1+\lambda_K{(4\pi\Lambda + 2\i\pi^2K -4\pi
K^2/\Lambda)}
\right)}.
\end{eqnarray}
\item[{Effective Range and Scattering Length}]
\label{sec:r0as}
The scattering wavefunction for incident momentum $\boldsymbol{K}$ and final momentum $\kbf$ is
\begin{eqnarray}
\tilde{a}_{\boldsymbol{K}}(\kbf) \equiv \delta(\boldsymbol{K} -\kbf)
-\frac{f(\kbf,\boldsymbol{K})}{2\pi^2(K^2-k^2+\i\eta)},
\end{eqnarray}
where $f(\kbf,\boldsymbol{K})$ is the scattering amplitude and is
thus given by
\begin{eqnarray}
\label{eq:f}
f(\kbf,\boldsymbol{K})=\frac{-2\pi^2\lambda_K}{1-\lambda_K\int_0^\Lambda{\rmd\kbf\,
\frac{1}{K^2-k^2+\i\eta}}}
= \left(-\frac{1}{2\pi^2\lambda_K} -  \frac{2}{\pi}\Lambda -
\i K + \frac{1}{\pi} K\ln\left( \frac{\Lambda + K}{\Lambda -
K}\right) \right)^{-1}.
\end{eqnarray}
The scattering amplitude for small incident momenta can be
approximated by the effective range expansion,
\begin{eqnarray}\label{ScattAmp}
f(K) &=& \frac{1}{K\cot\delta - \i K} \approx \frac{1}{-{1/a_s} - \i
K+ r_0K^2/2},
\end{eqnarray}
where $\delta$ is the phase shift, $a_s$ is the s-wave scattering
length and $r_0$ is the
effective range of the potential. We expand (\ref{eq:f}) for small $K$ using the expansions
\begin{eqnarray}
\label{eq:lambdaExp}
\frac{1}{\lambda_K} &\approx& \frac{(2\pi)^3\hbar^2}{m}\left(U_{aa} -
\frac{g^2}{2\varepsilon} \right)^{-1} +
\pi K^2\left(\frac{4\pi\hbar^2}{m}\frac{g^2/2
\varepsilon}{g(U_{aa}-g^2/2
\varepsilon)}\right)^2,\\
\label{eq:logExp}
\ln\left( \frac{\Lambda + K}{\Lambda - K}\right) &\approx&
\frac{2K}{\Lambda},
\end{eqnarray}
where (\ref{eq:logExp}) is valid as long as 
\begin{eqnarray}\label{valid1}
K\ll\Lambda,
\end{eqnarray}
and (\ref{eq:lambdaExp}) is valid as long as
\begin{eqnarray}\label{valid2}
\frac{\hbar^2K^2}{m}\ll \frac{g^2}{2U_{aa}}-\varepsilon .
\end{eqnarray}
We can thus use equations  (\ref{eq:f}) and (\ref{ScattAmp}) to
identify
\begin{eqnarray}
\label{eq:a_from_f}
a_s &=&\left[\frac{4\pi\hbar^2}{m}\left(U_{aa} -
\frac{g^2}{2\varepsilon} \right)^{-1}  +
\frac{2}{\pi}\Lambda\right]^{-1},
 \\
\label{eq:r0}
r_0 &=& \frac{4}{\pi \Lambda} -
\frac{1}{\pi}\left(\frac{4\pi\hbar^2}{m}\frac{g/2
\varepsilon}{(U_{aa}-g^2/2
\varepsilon)}\right)^2.
\end{eqnarray}

\item[Background Scattering Length]
Setting $ g^2/\varepsilon \to 0$ in (\ref{eq:a_from_f}) corresponds 
to the case when the effect of the Feshbach resonance is negligible;
ie, only the background scattering term is nonzero.  
This means that the background scattering length is
\begin{eqnarray}\label{bgsl}
a_{bg} &=& \left[{4\pi\hbar^2\over m U_{aa}}+{2\over
\pi}\Lambda\right].
\end{eqnarray}

\end{description}

\subsubsection{Nature of Effective Range Expansion}
\label{sec:NatureEffRange}%
\begin{figure}[t]
\begin{center}\large
\psfrag{s05}[t][t]{\color[rgb]{0,0,0}\setlength{\tabcolsep}{0pt}\begin{tabular}{c}$\text{Re}(K)/\Lambda$\end{tabular}}%
\psfrag{s06}[b][b]{\color[rgb]{0,0,0}\setlength{\tabcolsep}{0pt}\begin{tabular}{c}$\text{Im}(K)/\Lambda$\end{tabular}}%
\psfrag{x01}[t][t]{0}%
\psfrag{x02}[t][t]{0.1}%
\psfrag{x03}[t][t]{0.2}%
\psfrag{x04}[t][t]{0.3}%
\psfrag{x05}[t][t]{0.4}%
\psfrag{x06}[t][t]{0.5}%
\psfrag{x07}[t][t]{0.6}%
\psfrag{x08}[t][t]{0.7}%
\psfrag{x09}[t][t]{0.8}%
\psfrag{x10}[t][t]{0.9}%
\psfrag{x11}[t][t]{1}%
\psfrag{x12}[t][t]{-15}%
\psfrag{x13}[t][t]{-10}%
\psfrag{x14}[t][t]{-5}%
\psfrag{x15}[t][t]{0}%
\psfrag{x16}[t][t]{5}%
\psfrag{x17}[t][t]{10}%
\psfrag{x18}[t][t]{15}%
\psfrag{v01}[r][r]{0}%
\psfrag{v02}[r][r]{0.1}%
\psfrag{v03}[r][r]{0.2}%
\psfrag{v04}[r][r]{0.3}%
\psfrag{v05}[r][r]{0.4}%
\psfrag{v06}[r][r]{0.5}%
\psfrag{v07}[r][r]{0.6}%
\psfrag{v08}[r][r]{0.7}%
\psfrag{v09}[r][r]{0.8}%
\psfrag{v10}[r][r]{0.9}%
\psfrag{v11}[r][r]{1}%
\psfrag{v12}[r][r]{-15}%
\psfrag{v13}[r][r]{-10}%
\psfrag{v14}[r][r]{-5}%
\psfrag{v15}[r][r]{0}%
\psfrag{v16}[r][r]{5}%
\psfrag{v17}[r][r]{10}%
\psfrag{v18}[r][r]{15}%
\resizebox{8cm}{!}{\includegraphics{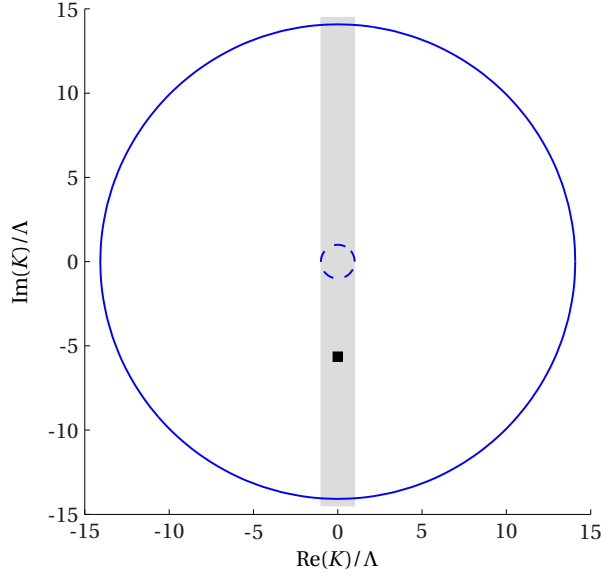}}%
\caption{The grey region shows the region of validity for the
scattering amplitude (\ref{eq:f}) in the complex plane. The upper
bounds (\ref{valid1}, \ref{valid2}) for the individual expansions
(\ref{eq:lambdaExp}, \ref{eq:logExp}) are given as the dashed and
solid circles respectively. The region of validity for the effective
range expansion is obviously situated within the smaller of these,
while the bound state solution (black square) can be found far
outside this area. The plot here is for a cutoff value of $\Lambda =
2.2\times 10^6$ and a scattering length of $a=900a_0$, but we get
qualitatively similar results for the whole parameter range used in
this paper.}
\label{fig:scatteringAmp}
\end{center}
\end{figure}

The effective range expansion of the scattering amplitude has the
rather limited range of validity---the two 
conditions (\ref{valid1}, \ref{valid2}) set an upper bound  for $ K$,
and for the expansion to be quantitatively valid, $ K$ must lie well
inside 
the region defined by the intersection of the spaces determined by
these bounds. The situation is illustrated in
Fig.\,\ref{fig:scatteringAmp} for parameters appropriate to the
problems we wish to study, for which available $K$-space is limited
by the bound determined by $\Lambda$, that is, by the dashed circle.

However, the exact solution (\ref{eq:f}) for the scattering amplitude
possesses singularities at $K=\pm\Lambda$, but is analytic  for
imaginary values of $K$ with modulus very much greater than
$\Lambda$.  In fact, the bound state solution to the scattering
problem is given by the pole of (\ref{eq:f}) on the negative
imaginary axis and, in cases we consider, is well outside the region bounded by the dashed
line.  This pole cannot be obtained by making the effective range
expansion 
(\ref{ScattAmp}) for the scattering amplitude. (In fact in
Section\,\ref{sec:boundstate}
we will find it more convenient to determine the bound state
properties by direct solution of the scattering equation, rather than
by direct evaluation of the position and residue of the pole, which
is fully equivalent to determining the pole and its residue).
We will therefore interpret the available $K$-space region as the
asymmetric region bound by the cutoff $\Lambda$ on the real axis, but
including the bound state on the imaginary axis, indicated by the
shaded area in Fig.\,\ref{fig:scatteringAmp}.

 With the model we have chosen, it is not possible to simultaneously
fit the binding energy, the scattering length and the effective
range, since there are only two fitting parameters, $g$ and
$\varepsilon$.
The scattering length and binding energy are the relevant measurable
parameters for the kind of problem we wish to study, whereas the
effective range does not play any direct role, and therefore we will
determine $g$ and $\varepsilon$ by fitting the measured values of
scattering length and binding energy.  In this paper, we will not
concern ourselves any further with the effective range.

\subsection{Bound state solution}
\label{sec:boundstate}%
In this atom-molecule model, a physical bound state will be a superposition of a
state of two $\psi$-field ``atoms'', and a state of one $\phi$-field
``molecule''.  A bound state solution is characterized by a negative
energy eigenvalue, $E\rightarrow -\hbar^2\alpha^2/m$ in
(\ref{eq:yamaguchi}), where $\alpha^2>0$, so that $K\rightarrow(0,\i\alpha)$.  Equation (\ref{eq:yamaguchi})
can now be written as
\begin{eqnarray}
\tilde\psi(\kbf) ={-}
\frac{\lambda_\alpha}{\alpha^2+k^2}\int_0^\Lambda{\rmd\kbf'\,\tilde{\psi}(
\kbf')},
\end{eqnarray}
where we have defined $\lambda_\alpha$ by setting $E\rightarrow
-\hbar^2\alpha^2/m$ in (\ref{eq:yamaguchi_coeff}).
By taking the integral over $\kbf$ on both sides, we can remove the
dependence on the wavefunction, to get
\begin{eqnarray}
\label{eq:bindingIntegral}
1 = -4\pi\int^\Lambda_0{\frac{\lambda_\alpha k^2}{\alpha^2+k^2}\rmd
k} = -4\pi\lambda_\alpha\left[\Lambda -
\alpha\arctan\left(\frac{\Lambda}{\alpha}\right)\right] , 
\end{eqnarray}
so that
$\alpha$, and hence the binding energy, are determined by the solution
of the equation
\begin{eqnarray}
\label{eq:boundstate}
\alpha\arctan\left(\frac{\Lambda}{\alpha}\right)= \left[\frac{
m}{2\pi^2\hbar^2}\left(U_{aa}+\frac{g^2/2}{-\hbar^2\alpha^2/m-\varepsilon}
\right)\right]^{-1}
+\Lambda. 
\end{eqnarray}

\begin{description}
\item[Threshold of binding]
At the bound state threshold, where $\alpha \rightarrow 0$, we can
solve (\ref{eq:boundstate}) exactly in terms of $\varepsilon$ to give
\begin{eqnarray}
\label{eq:threshold}
\varepsilon_\text{threshold} = \frac{g^2}{2}\left[U_{aa} +
\frac{2\pi^2\hbar^2}{m\Lambda}\right]^{-1},
\end{eqnarray}
which is of course identical to the solution of the scattering length
equation
(\ref{eq:a_from_f}) when $ a_s=0$.  Substituting from (\ref{bgsl}),
this takes 
the form
\begin{eqnarray}\label{eq:threshold2}
\varepsilon_\text{threshold} &=&
\left({g^2\over 2}\right)
\left({m\over 4\pi\hbar^2}\right)
{\pi/2\Lambda\over 1 - 2\Lambda a_{bg}/\pi}\,.
\end{eqnarray}
This is nonzero unless $ g^2\to 0$ at threshold.  
\item[{Weakly bound molecules}]
For small values of $\alpha$ we can expand the left hand side of
(\ref{eq:boundstate}) to give
\begin{equation}
\alpha\left(\frac{\pi}{2} - \frac{\alpha}{\Lambda} \right) =
\left[\frac{
m}{2\pi^2\hbar^2}\left(U_{aa}+\frac{g^2/2}{-\hbar^2\alpha^2/m-\varepsilon}
\right)\right]^{-1}
+\Lambda, 
\end{equation}
so that we get for weakly bound molecules
\begin{equation}
\label{eq:kappaS}
\alpha \approx \left[ \frac{
m}{4\pi\hbar^2}\left(U_{aa}+\frac{g^2/2}{-\varepsilon}\right)\right]^{-1}
+\frac{2}{\pi}\Lambda.
\end{equation}
It is clear that in the case of weakly bound molecules $\alpha
\approx 1/a_s$, as expected from the relationship between the binding
energy and the scattering length for sufficiently small binding
energies.

\item[{Tightly bound molecules}]
For large values of $\alpha$, expanding $\arctan (\Lambda/\alpha)$ in
(\ref{eq:boundstate}) yields
\begin{eqnarray}
\alpha\left(\frac{\Lambda}{\alpha} - \frac{\Lambda^3}{3\alpha^3}
\right) \approx \left[\frac{
m}{2\pi^2\hbar^2}\left(U_{aa}-\frac{g^2/2}{\hbar^2\alpha^2/m+\varepsilon}
\right)\right]^{-1}
+\Lambda, 
\end{eqnarray}
so that we instead get
\begin{eqnarray}\label{epslarge}
    \alpha^{2}&\approx&- {m\varepsilon\over\hbar^{2}}
    +
    {\pi^{2}\hbar^{2}\Lambda^{3}g^{2}\over 3m}
   \left( {2\pi^{2}\Lambda^{3}U_{aa}\over 3} +\varepsilon\right)^{-1}.
\end{eqnarray}
For sufficiently large values of $\varepsilon$ we have that $\alpha^2
\approx -m\varepsilon/\hbar^2$, which means that the detuning $\varepsilon$
is approximately equal to the binding energy $E_b =
\hbar^2\alpha^2/m$ for large values of the detuning, as would be
expected.
\end{description}
\subsubsection{Atomic Fraction of the Bound State}
The bound state solution for the atom wavefunction is
\begin{eqnarray}
\label{eq:atomsol}
\tilde{\psi}(\kbf) = \frac{N}{\alpha^2+k^2},
\end{eqnarray}
where, if the normalization is chosen to be
 $\int d^3\vec k\,|\tilde{\psi}(\kbf)|^2=M$, the factor $N$ will be
given by
\begin{eqnarray}
\fl
\frac{M}{N^2} = \int^\Lambda_0{\rmd\kbf\,
\frac{1}{(\alpha^2+k^2)^2}}=4\pi\int^\Lambda_0{\frac{k^2}{(\alpha^2+k^2)^2}dk}
= 2\pi\left[
\frac{1}{\alpha}\arctan\left(\frac{\Lambda}{\alpha}\right) 
-\frac{\Lambda}{\Lambda^2+\alpha^2}
\right] .
\end{eqnarray}
Similarly the solution for the molecular state can be obtained by
substituting (\ref{eq:atomsol}) into (\ref{eq:molyamaguchi}), giving
\begin{eqnarray}\label{phinorm1}
\fl
\tilde{\phi}(0) &=&
 -\frac{g}{2\left(\hbar^2\alpha^2/m +\varepsilon\right)}
    \int_0^\Lambda{\rmd\kbf'\,\tilde{\psi}(\kbf')},
\\ 
\label{phinorm2} &=&
-\frac{2\pi gN}{\hbar^2\alpha^2/m + \varepsilon}\left[\Lambda -
\alpha\arctan\left(\frac{\Lambda}{\alpha}\right)\right],
\\ 
\label{phinorm3}
&=&  
{4\pi N\over g}\left(U_{aa}
\left(\alpha\arctan\left({\Lambda\over\alpha}\right) -\Lambda\right)  
-{2\pi^2\hbar^2\over m}\right),
\end{eqnarray}
where we have used (\ref{eq:boundstate}) to get to the last line.

\begin{figure}[t]
\begin{center}
\begin{psfrags}%
{\footnotesize
\psfrag{s01}[t][t]{\color[rgb]{0,0,0}\setlength{\tabcolsep}{0pt}\begin{tabular}{c}$\alpha/\Lambda$\end{tabular}}%
\psfrag{s02}[b][b]{\color[rgb]{0,0,0}\setlength{\tabcolsep}{0pt}\begin{tabular}{c}Fraction\end{tabular}}%
\psfrag{x01}[t][t]{0}%
\psfrag{x02}[t][t]{0.2}%
\psfrag{x03}[t][t]{0.4}%
\psfrag{x04}[t][t]{0.6}%
\psfrag{x05}[t][t]{0.8}%
\psfrag{x06}[t][t]{1}%
\psfrag{x07}[t][t]{$10^{-5}$}%
\psfrag{x08}[t][t]{$10^{0}$}%
\psfrag{v01}[r][r]{0}%
\psfrag{v02}[r][r]{0.2}%
\psfrag{v03}[r][r]{0.4}%
\psfrag{v04}[r][r]{0.6}%
\psfrag{v05}[r][r]{0.8}%
\psfrag{v06}[r][r]{1}%
\psfrag{v07}[r][r]{0}%
\psfrag{v08}[r][r]{0.5}%
\psfrag{v09}[r][r]{1}%
\resizebox{10cm}{!}
{\includegraphics{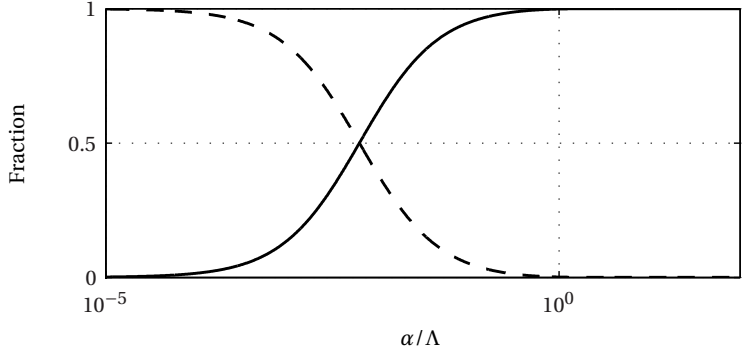}}}%
\end{psfrags}%
\caption{The fractions of the atom pairs (dashed line) and
elementary molecules (solid line).  The values of the length
parameters are in this case those for $^{85}$Rb; namely $ l_{ug}
\approx 10^{-7}$, $a_{bg}=-2.32\times 10^{-8}$, and we have chosen
{$\Lambda = 2\times10^6$}.}
\label{fig:ratios}
\end{center}
\end{figure}

The ratio of the probabilities for the atom and the molecule states
can then be 
written
\begin{eqnarray}\label{eq:ratio}
\frac{P_{\text{mol}}}{P_{\text{atm}}}
&=&
{|\phi(0)|^2\over \int \rmd\vec k\,|\tilde{\psi}(\kbf)|^2}
=
8\pi\left(\alpha\over\Lambda\right)^2(l_{ug}\Lambda)^3
\left(
\left( t(\alpha/\Lambda) -\pi/2\Lambda a_{bg}\right)^2\over
  t(\alpha/\Lambda) +\Lambda^2/(\Lambda^2+\alpha^2)
\right),
\end{eqnarray}
where\begin{eqnarray}
\label{tx}
 t(z) &\equiv & z\arctan(1/z)-1,
\\ \label{lug}
l_{ug} &\equiv& \left(U_{aa}\over g\right)^{2/3}.
\end{eqnarray}
The result depends on the four parameters which are all lengths,
namely $ l_{ug}$, $ a_{bg}$, $ \Lambda^{-1}$ and $ \alpha^{-1}$.   
It is worth noting that while it is mandatory that $|K|/\Lambda$ be
less than 
1, the same is not true for $|\alpha |/\Lambda$, which does not
necessarily 
have to be less than  1. On the contrary, in most experiments the
binding energy is usually such that $|\alpha | \gg \Lambda$, for
realistic values of the cutoff (cf. Section
\ref{sec:NatureEffRange}). 

The parameter $ \alpha^{-1}$ determines the spatial extent of the
atom-pair part of the wavefunction of the bound state, and this can
only be represented on the spatial grid corresponding to the cutoff $
\Lambda$ if $ \alpha/\Lambda \ll 1$.  The crossover from the
predominantly atom-pair wavefunction to the elementary molecule
wavefunction corresponds to the molecular size becoming smaller than
the size of the spatial grid.  However, (\ref{eq:ratio}) shows that,
as well as $ \Lambda^{-1}$, there are the other lengths $ a_{bg}$ and
$ l_{ug}$ which also come into play.  This means that the intuitive
idea that the crossover happens at $ \alpha/\Lambda\approx 1$ is
significantly modified, and in Fig.\,\ref{fig:ratios}, which shows the
fractions of the atom and the molecule states as functions of
$\alpha/\Lambda$, it can be seen that the crossover is closer to $
\alpha/\Lambda\approx 0.01$.

The crossover is thus explicitly cutoff dependent, but this is purely
a technical issue.  There is no fundamental physics in this crossover;
it is determined simply by the technical need to have a cutoff.  This
has to be sufficiently large to represent the spatial features under
investigation, and also has an upper bound determined by the
requirement that $ \Lambda a_s$ be much less than one.  However,
within this range it is arbitrary.

\subsection{Determination of Parameters}
\label{sec:parameters}%
Of the parameters in the Hamiltonian (\ref{eq:H}) necessary for our
modelling, 
$g$, $\varepsilon$ and $U_{aa}$, it is only the last that can be
directly 
determined experimentally, since the background 
scattering length $a_{bg}$ is given by (\ref{bgsl}).
The coupling $g$ and the energy detuning
$\varepsilon$, however,  need to be derived from other physically
measurable
quantities. 

\begin{figure}
\begin{center}
\begin{psfrags}
\psfrag{s02}[t][t]{\color[rgb]{0,0,0}\setlength{\tabcolsep}{0pt}\begin{tabular}{c}$1/a_s$
$[a_0^{-1}]$\end{tabular}}%
\psfrag{s03}[b][b]{\color[rgb]{0,0,0}\setlength{\tabcolsep}{0pt}\begin{tabular}{c}$\varepsilon$
[MHz]\end{tabular}}%
\psfrag{s06}[t][t]{\color[rgb]{0,0,0}\setlength{\tabcolsep}{0pt}\begin{tabular}{c}$1/a_s$
$[a_0^{-1}]$\end{tabular}}%
\psfrag{s07}[b][b]{\color[rgb]{0,0,0}\setlength{\tabcolsep}{0pt}\begin{tabular}{c}$g$
[Jm$^{-3/2}$]\end{tabular}}%
\psfrag{s09}[lt][lt]{\color[rgb]{0,0,0}\setlength{\tabcolsep}{0pt}\begin{tabular}{l}$\times10^{-3}$\end{tabular}}%
\psfrag{s14}[lt][lt]{\color[rgb]{0,0,0}\setlength{\tabcolsep}{0pt}\begin{tabular}{l}$\times10^{-3}$\end{tabular}}%
\psfrag{s15}[lt][lt]{\color[rgb]{0,0,0}\setlength{\tabcolsep}{0pt}\begin{tabular}{l}$\times10^{-39}$\end{tabular}}%
\psfrag{h1}[lt][lt]{\color[rgb]{0,0,0}\setlength{\tabcolsep}{0pt}\begin{tabular}{l}(a)\end{tabular}}%
\psfrag{h2}[lt][lt]{\color[rgb]{0,0,0}\setlength{\tabcolsep}{0pt}\begin{tabular}{l}(b)\end{tabular}}%
\psfrag{x01}[t][t]{0}%
\psfrag{x02}[t][t]{0.1}%
\psfrag{x03}[t][t]{0.2}%
\psfrag{x04}[t][t]{0.3}%
\psfrag{x05}[t][t]{0.4}%
\psfrag{x06}[t][t]{0.5}%
\psfrag{x07}[t][t]{0.6}%
\psfrag{x08}[t][t]{0.7}%
\psfrag{x09}[t][t]{0.8}%
\psfrag{x10}[t][t]{0.9}%
\psfrag{x11}[t][t]{1}%
\psfrag{x12}[t][t]{0}%
\psfrag{x13}[t][t]{2}%
\psfrag{x14}[t][t]{4}%
\psfrag{x15}[t][t]{0}%
\psfrag{x16}[t][t]{2}%
\psfrag{x17}[t][t]{4}%
\psfrag{v01}[r][r]{0}%
\psfrag{v02}[r][r]{0.1}%
\psfrag{v03}[r][r]{0.2}%
\psfrag{v04}[r][r]{0.3}%
\psfrag{v05}[r][r]{0.4}%
\psfrag{v06}[r][r]{0.5}%
\psfrag{v07}[r][r]{0.6}%
\psfrag{v08}[r][r]{0.7}%
\psfrag{v09}[r][r]{0.8}%
\psfrag{v10}[r][r]{0.9}%
\psfrag{v11}[r][r]{1}%
\psfrag{v12}[r][r]{1}%
\psfrag{v13}[r][r]{2}%
\psfrag{v14}[r][r]{3}%
\psfrag{v15}[r][r]{4}%
\psfrag{v16}[r][r]{5}%
\psfrag{v17}[r][r]{6}%
\psfrag{v18}[r][r]{7}%
\psfrag{v19}[r][r]{8}%
\psfrag{v20}[r][r]{-1}%
\psfrag{v21}[r][r]{-0.8}%
\psfrag{v22}[r][r]{-0.6}%
\psfrag{v23}[r][r]{-0.4}%
\psfrag{v24}[r][r]{-0.2}%
\psfrag{v25}[r][r]{0}%
\resizebox{12cm}{!}{\includegraphics{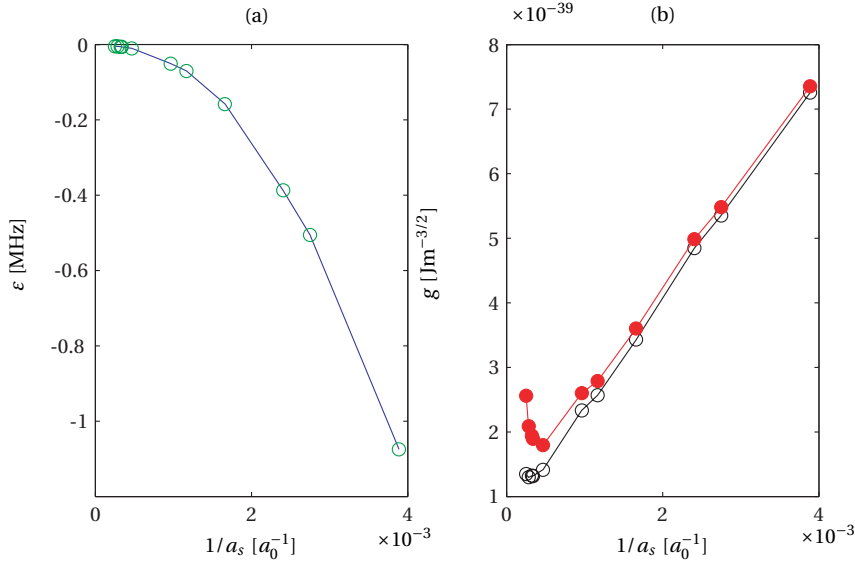}}%
\end{psfrags}%
\caption{Values in SI units of the parameters a) $g$, and b) $\varepsilon$
in the Hamiltonian (\ref{eq:H}), as calculated from
(\ref{eq:a_from_f}) and (\ref{eq:boundstate}) for cutoff values
$\Lambda = 2.2\times 10^{6}\text{m}^{-1}$ (hollow black circles) and
$\Lambda = 1\times 10^{7}\text{m}^{-1}$ (solid red circles).  For (a) only values of $ \varepsilon$ corresponding to $
\Lambda=2.2\times10^6$ are plotted, since the values of $ \varepsilon$
for the two different values $ \Lambda$ are almost indistinguishable
from each other.  In both plots the parameters are calculated using
data of the binding energy and scattering length taken from
\cite{claussen2003}. Solid lines are to guide the eye.}
\label{g-eps}
\end{center}
\end{figure}

The s-wave scattering length $a_s$ is well known for most condensate
systems, and for $ ^{85}\text{Rb}$ the molecular binding energy $E_b
\equiv \hbar^2\alpha^2/m$ has been measured for a range of magnetic
fields which covers that used to vary the scattering length by means
of a Feshbach resonance.  Thus, a range of corresponding experimental
values of $ a_s,\alpha$ exists, and matching these to the expressions
(\ref{eq:a_from_f}) and (\ref{eq:boundstate}), it is possible to
determine the necessary sets of values of $g$ and $\varepsilon$ for a
chosen value of the cutoff $\Lambda$ in the form
\begin{eqnarray}
\label{eq:epsilon}
\varepsilon &=&\left({\hbar^2\alpha^2\over 2m}\right) 
\frac{\big(\pi-2\Lambda a_s\big)
\big (1-2\Lambda t(\alpha/\Lambda) a_{bg}/\pi\big)}
{\Lambda a_s(1+t(\alpha/\Lambda))-\pi} 
\,,\\
\label{eq:g2}
g^2
&=&\left( {8\pi\hbar^4\alpha^2\over m^2}\right)
{\big(a_{bg}(\pi-2\Lambda a_s) -\pi a_s\big)
\big(1-2\Lambda t(\alpha/\Lambda)a_{bg}/\pi\big)
\over
2\Lambda a_s \big(1+ t(\alpha/\Lambda\big) -\pi}\, .
\end{eqnarray}
In Fig.\,\ref{g-eps} we show the Hamiltonian parameters as functions
of the s-wave scattering length calculated in this way for two
different values of the cutoff $\Lambda$, using data of the
binding energy and scattering length taken from \cite{claussen2003}.
The dependence on $ \Lambda$ is seen to be quite weak. (In
applications to c-field calculations, the cutoff is related to the
size of the simulation grid and it is important to choose the grid
such that $a_s$ and $E_b$ depend only weakly on the value of
$\Lambda$.)

\begin{figure}[t]
\begin{center}
\begin{psfrags}
\psfrag{s01}[t][t]{\color[rgb]{0,0,0}\setlength{\tabcolsep}{0pt}\begin{tabular}{c}$1/a_s$
$[a_0^{-1}]$\end{tabular}}%
\psfrag{s02}[b][b]{\color[rgb]{0,0,0}\setlength{\tabcolsep}{0pt}\begin{tabular}{c}$E_b/\varepsilon$\end{tabular}}%
\psfrag{s05}[t][t]{\color[rgb]{0,0,0}\setlength{\tabcolsep}{0pt}\begin{tabular}{c}$1/a_s$
$[a_0^{-1}]$\end{tabular}}%
\psfrag{s06}[b][b]{\color[rgb]{0,0,0}\setlength{\tabcolsep}{0pt}\begin{tabular}{c}$\alpha$
$[a_0^{-1}]$\end{tabular}}%
\psfrag{s13}[lt][lt]{\color[rgb]{0,0,0}\setlength{\tabcolsep}{0pt}\begin{tabular}{l}$\times10^{-3}$\end{tabular}}%
\psfrag{s14}[lt][lt]{\color[rgb]{0,0,0}\setlength{\tabcolsep}{0pt}\begin{tabular}{l}$\times10^{-3}$\end{tabular}}%
\psfrag{s15}[lt][lt]{\color[rgb]{0,0,0}\setlength{\tabcolsep}{0pt}\begin{tabular}{l}$\times10^{-3}$\end{tabular}}%
\psfrag{h1}[lt][lt]{\color[rgb]{0,0,0}\setlength{\tabcolsep}{0pt}\begin{tabular}{l}(a)\end{tabular}}%
\psfrag{h2}[lt][lt]{\color[rgb]{0,0,0}\setlength{\tabcolsep}{0pt}\begin{tabular}{l}(b)\end{tabular}}%
\psfrag{x01}[t][t]{0}%
\psfrag{x02}[t][t]{0.1}%
\psfrag{x03}[t][t]{0.2}%
\psfrag{x04}[t][t]{0.3}%
\psfrag{x05}[t][t]{0.4}%
\psfrag{x06}[t][t]{0.5}%
\psfrag{x07}[t][t]{0.6}%
\psfrag{x08}[t][t]{0.7}%
\psfrag{x09}[t][t]{0.8}%
\psfrag{x10}[t][t]{0.9}%
\psfrag{x11}[t][t]{1}%
\psfrag{x12}[t][t]{0}%
\psfrag{x13}[t][t]{1}%
\psfrag{x14}[t][t]{2}%
\psfrag{x15}[t][t]{3}%
\psfrag{x16}[t][t]{0}%
\psfrag{x17}[t][t]{2}%
\psfrag{x18}[t][t]{4}%
\psfrag{v01}[r][r]{0}%
\psfrag{v02}[r][r]{0.1}%
\psfrag{v03}[r][r]{0.2}%
\psfrag{v04}[r][r]{0.3}%
\psfrag{v05}[r][r]{0.4}%
\psfrag{v06}[r][r]{0.5}%
\psfrag{v07}[r][r]{0.6}%
\psfrag{v08}[r][r]{0.7}%
\psfrag{v09}[r][r]{0.8}%
\psfrag{v10}[r][r]{0.9}%
\psfrag{v11}[r][r]{1}%
\psfrag{v12}[r][r]{0}%
\psfrag{v13}[r][r]{0.5}%
\psfrag{v14}[r][r]{1}%
\psfrag{v15}[r][r]{1.5}%
\psfrag{v16}[r][r]{2}%
\psfrag{v17}[r][r]{2.5}%
\psfrag{v18}[r][r]{3}%
\psfrag{v19}[r][r]{1}%
\psfrag{v20}[r][r]{1.02}%
\psfrag{v21}[r][r]{1.04}%
\psfrag{v22}[r][r]{1.06}%
\psfrag{v23}[r][r]{1.08}%
\psfrag{v24}[r][r]{1.1}%
\resizebox{12cm}{!}{\includegraphics{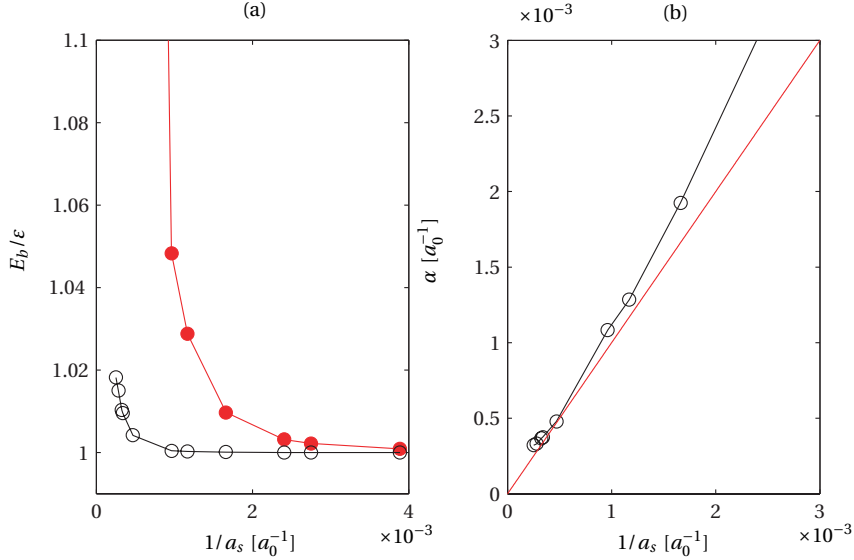}}%
\end{psfrags}%
\caption{a) The ratio $ E_b/\varepsilon$ plotted for cutoff values
$\Lambda = 2.2\times 10^{6}\text{m}^{-1}$ (hollow black circles) and
$\Lambda = 1\times 10^{7}\text{m}^{-1}$ (solid red circles).  This
ratio is seen to be very close to 1, deviating only at scattering
lengths more than of about $5\times 10^{-8}\text{m}\approx
1000a_\text{Bohr} $.  The deviation at larger scattering lengths
depends strongly on the choice of cutoff $ \Lambda$, and is also
affected by mean field effects; 
b) Plot of $\alpha$  vs. $ a_s^{-1}$; the value of $ \alpha$ must
approach $ a_s^{-1}$ for sufficiently small $ \alpha$.  However, the
data used for these plots is measured in a \BEC, and the four points
corresponding to $ a_s^{-1}< 1\times 10^7\text{m}^{-1}$ deviate from
this law because of mean field effects specific to that situation.  In
both plots the parameters are calculated using data of the binding
energy and scattering length taken from \cite{claussen2003}.  }
\label{parms-analysis}
\end{center}
\end{figure}

Figs.\,\ref{g-eps} and \ref{parms-analysis} also show that the parameter $g$ is essentially a
linear function of $a_s^{-1}$, while energy offset $ \varepsilon$ is
almost proportional to the binding energy $ E_b$.  Looking at
Fig.\,\ref{parms-analysis}, it can be seen that $E_b$ and
$\varepsilon$ are essentially equal except for the very small binding
energies which occur at large $ a_s$.  In fact, since the data show
that $ g$ is unlikely to be zero at the threshold of binding, where $
a_s^{-1} \to 0$, it can be seen from (\ref{eq:threshold}) that $
\varepsilon_{\text{threshold}} \ne 0$, and given that for $
^{85}\text{Rb}$ the background scattering length is negative, the
threshold value of $ \varepsilon $ is in fact positive.

In panel b) of Fig.\,\ref{parms-analysis}, it can also be seen that $
a_s^{-1}$
and $ \alpha$ become very close, but when equality approaches at weak 
binding, the mean field effect of the \BEC\ used in the experimental 
measurement obscures the equality, which was also noted in
\cite{claussen2003}.

\subsection{Validity of the Effective range Expansion}
\begin{figure}[t]
\begin{center}\footnotesize
\psfrag{s01}[t][t]{\color[rgb]{0,0,0}\setlength{\tabcolsep}{0pt}\begin{tabular}{c}$K/\Lambda$\end{tabular}}%
\psfrag{s02}[b][b]{\color[rgb]{0,0,0}\setlength{\tabcolsep}{0pt}\begin{tabular}{c}Ratio\end{tabular}}%
\psfrag{x01}[t][t]{0}%
\psfrag{x02}[t][t]{0.1}%
\psfrag{x03}[t][t]{0.2}%
\psfrag{x04}[t][t]{0.3}%
\psfrag{x05}[t][t]{0.4}%
\psfrag{x06}[t][t]{0.5}%
\psfrag{x07}[t][t]{0.6}%
\psfrag{x08}[t][t]{0.7}%
\psfrag{x09}[t][t]{0.8}%
\psfrag{x10}[t][t]{0.9}%
\psfrag{x11}[t][t]{1}%
\psfrag{x12}[t][t]{0}%
\psfrag{x13}[t][t]{0.2}%
\psfrag{x14}[t][t]{0.4}%
\psfrag{x15}[t][t]{0.6}%
\psfrag{x16}[t][t]{0.8}%
\psfrag{x17}[t][t]{1}%
\psfrag{v01}[r][r]{0}%
\psfrag{v02}[r][r]{0.1}%
\psfrag{v03}[r][r]{0.2}%
\psfrag{v04}[r][r]{0.3}%
\psfrag{v05}[r][r]{0.4}%
\psfrag{v06}[r][r]{0.5}%
\psfrag{v07}[r][r]{0.6}%
\psfrag{v08}[r][r]{0.7}%
\psfrag{v09}[r][r]{0.8}%
\psfrag{v10}[r][r]{0.9}%
\psfrag{v11}[r][r]{1}%
\psfrag{v12}[r][r]{0}%
\psfrag{v13}[r][r]{0.5}%
\psfrag{v14}[r][r]{1}%
\psfrag{v15}[r][r]{1.5}%
\psfrag{v16}[r][r]{2}%
{\includegraphics{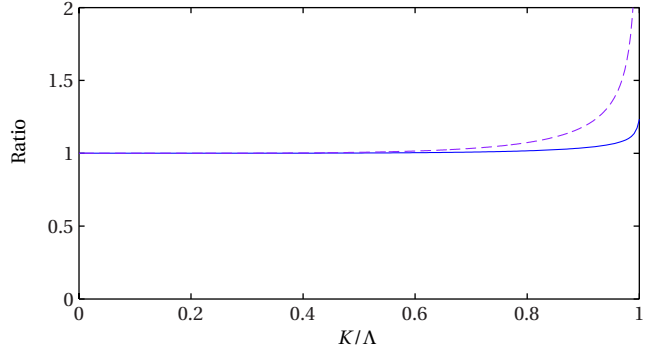}}%
\caption{Ratio of the effective range expansion and the exact
expression for $K\cot\delta$. The dash-dotted purple line corresponds
to $\Lambda=10^7$ and the blue solid line corresponds to $\Lambda =
2.2\times 10^6$. The scattering length is set to $a_s = 900a_0$ and
the effective range is given by equation (\ref{eq:r0}). }
\label{fig:kcot}
\end{center}
\end{figure}

Using (\ref{eq:epsilon}) and (\ref{eq:g2}) we can now estimate the
validity of the effective range expansion in section \ref{sec:r0as}.
In Fig.\,\ref{fig:kcot} we plot the ratio of the effective range
expansion $K\cot\delta\approx 1/a_s + r_0K^2/2$ and the exact
expression for $K\cot\delta$, given by the inverse of the real part
of the right hand side of (\ref{eq:f}). We can easily see that the
effective range expansion is only valid up to $K\approx 0.6\Lambda$.

\section{Conclusion}
The Hamiltonian (\ref{eq:H}) can thus be used to simulate realistic
atom-molecule systems, as long as the \emph{binding energy and
s-wave scattering length are known.  It is not important to match the
value of the \emph{effective range}, which in this kind of model is
determined by the value of the cutoff, and has little physical
significance.} The mapping of these physically observable
parameters to the phenomenological parameters in the Hamiltonian will
be dependent on the choice of the cutoff, but the results obtained in
simulations should be essentially independent of the choice of
$\Lambda$.

In \paperii\ and \paperiii, we will apply the model Hamiltonian we
have formulated here to the application of c-field methods to:
\begin{enumerate}
\item The formulation (for this model) of the simplest mean-field
theory, and in particular the Thomas-Fermi approximation to the profile
of a trapped condensate.
    
\item 
The study of the excitation spectrum of \BEC s and application to
Bragg scattering of a homogeneous system;

\item The modelling of
Feshbach-resonance-enhanced Bragg scattering from a trapped
inhomogeneous \BEC.
\end{enumerate}
We will demonstrate that the experimental data from \cite{papp2008}
can be reproduced accurately.

\subsubsection*{Acknowledgments} The research in this paper was
supported by 
the New Zealand Foundation for Research, Science and
Technology under Contract No. NERF-UOOX0703, ``Quantum
Technologies'', Marsden Contract No. UOO509.

\appendix
\section{Corrections to the Scattering Length and Binding Energy for Non-Zero Centre of Mass Momentum}

In Section \ref{sec:2atom1mol} we assume that the centre of mass momentum of the system is zero, and thus we solve the integral in (\ref{eq:yamaguchi}) on the range $[0,\Lambda]$. In this appendix we are investigating the corrections that occur when the centre of mass momentum is non-zero.

\subsection{Scattering length}
To find the corrections to the scattering length, we wish to evaluate the integral in (\ref{eq:scatteringPsi})
\begin{eqnarray}
{\mathcal I}_s = \int_{{\cal R}_{\vec Q}}{\rmd{\bf q}\,\frac{1}{K^2-q^2+\i\eta}} 
\end{eqnarray}
where the range $ {\cal R}_{\vec Q}$ is now defined by
\begin{eqnarray}\label{ap2}
\left | {\bf Q} \pm2{\bf q} \right | &\le & 2\Lambda,
\end{eqnarray}
for the centre of mass momentum
 $ \hbar{\bf Q}\equiv {\bf p}_1+{\bf p}_2 = {\bf p}_3+{\bf p}_4$.

We change the integral to polar coordinates, so that the range $ {{\cal R}_{\vec Q}}$ 
is equivalent to
\begin{eqnarray}\label{ap6}
 Q^2 + 4{{q}}^2 \pm 4 {q}Q\cos\theta \le 4\Lambda^2
\\ \label{ap7}
\Longrightarrow   |\cos\theta | 
 \le X\equiv \min\left(1, { 4\Lambda^2 -   Q^2 - 4{{q}}^2 \over 4  {q}Q} 
\right).
\end{eqnarray}
Thus, we can write
\begin{eqnarray}\label{ap8}
{\cal I}_s  &=& 
2\pi\int_0^{\sqrt{\Lambda^2-Q^2/4}}\frac{q^2}{K^2-q^2+\i\eta}d{q} \int_0^{X}2d\cos\theta 
\\ \label{ap9}
&=& 4\pi  \int_0^{\sqrt{\Lambda^2-Q^2/4}}\frac{q^2}{K^2-q^2+\i\eta}d{q}\,
\min\left(1, { 4\Lambda^2 -   Q^2 - 4{{q}}^2 \over 4  {q}Q}\right).
\nonumber\\
\end{eqnarray}
Noting that the point at which 
$ {\left( 4\Lambda^2 -   Q^2 - 4{{q}}^2\right) / 4 m {q}Q}=1$ is when
$ {q} = 2\Lambda - Q$, we can write
\begin{eqnarray}\fl\label{ap10}
{\cal I}_s  &=& 
 4\pi  \int_{\Lambda-Q/2}^{\sqrt{\Lambda^2-Q^2/4}}\,\frac{q^2}{K^2-q^2+\i\eta}
{ 4\Lambda^2 -   Q^2 - 4{{q}}^2 \over 4  {q}Q}\,d{q}
\nonumber\\
&&+ 4\pi  \int_{0}^{\Lambda-Q/2}\frac{q^2}{K^2-q^2+\i\eta}\,d{q} , \\
&=& 4\pi \left(\frac{\Lambda}{2}-\frac{Q}{4}+\frac{\Lambda^2-Q^2/4-K^2/2}{2Q}\ln\left|\frac{K^2-\left(\Lambda-Q/2\right)^2}{K^2-\Lambda^2+Q^2/4}\right|\right) \nonumber\\
&& + 4\pi \left(-\Lambda+\frac{Q}{2}-\frac{\i\pi}{2}K+\frac{K}{2}\ln\left|\frac{K+\Lambda-Q/2}{K-\Lambda+Q/2} \right|\right)
\\
\label{eq:integralResult}
&=& -2\pi\Lambda +\pi Q -2\i\pi^2K +2\pi\frac{\Lambda^2-Q^2/4-K^2/2}{Q}\ln\left|\frac{K^2-\left(\Lambda-Q/2\right)^2}{K^2-\Lambda^2+Q^2/4}\right| \nonumber \\
&&+2\pi K\ln\left|\frac{K+\Lambda-Q/2}{K-\Lambda+Q/2} \right| .
\end{eqnarray}

In the case when $K\ll\Lambda$, we can set $K\rightarrow0$ in (\ref{eq:integralResult}) and the expression for the scattering amplitude becomes
\begin{equation}
f=\frac{-2\pi^2\lambda_K}{1-\lambda_K{\mathcal I}_s}
= \left(-\frac{1}{2\pi^2\lambda_{K}} -  \frac{\Lambda}{\pi} +\frac{Q}{2\pi}+ \frac{1}{\pi}\frac{\Lambda^2-Q^2/4}{Q}\ln\left|\frac{\Lambda-Q/2}{\Lambda+Q/2}\right|\right)^{-1},
\end{equation}
giving us the following expression for the scattering length,
\begin{equation}
a_s = \left[\frac{4\pi\hbar^2}{m}\left(U_{aa} -
\frac{g^2}{2\varepsilon} \right)^{-1} +   \frac{\Lambda}{\pi} -\frac{Q}{2\pi}- \frac{1}{\pi}\frac{\Lambda^2-Q^2/4}{Q}\ln\left|\frac{\Lambda-Q/2}{\Lambda+Q/2}\right| \right]^{-1}.
\end{equation}
For $Q\ll 2\Lambda$, this can be approximated by
\begin{equation}
a_s \approx \left[\frac{4\pi\hbar^2}{m}\left(U_{aa} -
\frac{g^2}{2\varepsilon} \right)^{-1} +  \frac{2}{\pi}\Lambda - \frac{Q}{2\pi} - \frac{Q^2}{6\pi\Lambda} \right]^{-1}.
\end{equation}
Thus, the consequence of a non-zero $Q$ is a reduction in the effective value of $\Lambda$. However, this reduction will have very little impact on the value of the scattering length, since in practice $\Lambda$ must be chosen to reduce any such effects.

\subsection{Binding Energy}
To find the correction to the binding energy, we wish to solve the integral in (\ref{eq:bindingIntegral})
\begin{eqnarray}
{\mathcal I}_b = \int_{{\cal R}_{\vec Q}}{\rmd{\bf q}\,\frac{1}{\alpha^2+q^2}} ,
\end{eqnarray}
where the range $ {\cal R}_{\vec Q}$ is now defined by
\begin{eqnarray}
\left | {\bf Q} \pm2{\bf q} \right | &\le & 2\Lambda,
\end{eqnarray}
for the centre of mass momentum
 $ \hbar{\bf Q}\equiv {\bf p}_1+{\bf p}_2 = {\bf p}_3+{\bf p}_4$.

Following the same method as above, we can write the integral as
\begin{eqnarray}\fl
{\cal I}_b  &=& 
 4\pi  \int_{\Lambda-Q/2}^{\sqrt{\Lambda^2-Q^2/4}}\,\frac{q^2}{\alpha^2+q^2}
{ 4\Lambda^2 -   Q^2 - 4{{q}}^2 \over 4  {q}Q}\,d{q} + 4\pi  \int_{0}^{\Lambda-Q/2}\frac{q^2}{\alpha^2+q^2}\,d{q} , \\
&=& \pi Q - 2\pi\Lambda + \pi \frac{4\Lambda^2-Q^2+4\alpha^2}{2Q}\ln\left({\frac{\alpha^2+\Lambda^2-Q^2/4}{\alpha^2+(\Lambda-Q/2)^2}}\right) \nonumber\\
&&+ 4\pi\Lambda -2\pi Q -4\pi\arctan{\left(\frac{\Lambda-Q/2}{\alpha}\right)}, \\
\label{eq:integralBEresult}
&\approx& - 4\pi\arctan{\left(\frac{\Lambda-Q/2}{\alpha}\right)} +4\pi\left(\Lambda-Q/2\right) + \frac{\pi Q}{\alpha^2+\Lambda^2}\left(\Lambda - Q/2\right)^2+ ... \,, 
\end{eqnarray} 
where the expansion on the last line is valid as long as $Q^2\ll4\alpha^2+4\Lambda^2$. This is a reasonable approximation, since $Q<2\Lambda$ always, and $\alpha\gg\Lambda$ usually holds. 
The expression for the binding energy corresponding to (\ref{eq:boundstate}) then becomes
\begin{equation}
\alpha\arctan\left(\frac{\Lambda-Q/2}{\alpha}\right) \approx \left[\frac{
m}{2\pi^2\hbar^2}\left(U_{aa}+\frac{g^2/2}{-\hbar^2\alpha^2/m-\varepsilon}
\right)\right]^{-1} 
+(\Lambda-Q/2)\left(1+
 \frac{Q\left(\Lambda - Q/2\right)}{4\left(\alpha^2+\Lambda^2\right)}
 \right). 
\end{equation}
The assumed condition  $Q^2\ll4\alpha^2+4\Lambda^2$ means that the final term on the right hand side is very little different from $\Lambda - Q/2$, so that the effect of non-zero $Q$ is to replace $\Lambda$ in the binding energy condition (\ref{eq:boundstate}) by $\Lambda - Q/2 < \Lambda$.  This reduction of $\Lambda$ will have very little effect on the binding energy, as can be seen in Fig.\,\ref{parms-analysis}.


\bibliographystyle{unsrt}
\bibliography{Bibliography}

\end{document}